\documentstyle[epsfig]{aipproc}

\begin{document}
\title{Ultra High Energy Cosmic Rays and Inflation.}

\author{Igor I. Tkachev}
\address{TH Division, CERN, CH-1211 Geneva 23, Switzerland\\
and\\
Institute for Nuclear Research, Russian Academy of Sciences,\\
60th October Anniversary Prosp. 7a, Moscow 117312, Russia}

%\lefthead{LEFT head}
%\righthead{RIGHT head}
\maketitle

\begin{abstract}
Two processes of matter creation after inflation: 1) gravitational creation
of superheavy (quasi)stable particles, and 2) non-thermal phase transitions
leading to formation of topological defects, may be relevant to 
the resolution of
the puzzle of cosmic rays observed with energies beyond GZK cut-off.
Both possibilities are reviewed in this talk.
\end{abstract}

\section*{Introduction}

According to the modern tale, all matter in the Universe was created
in reheating after inflation. While this happened really long ago
and on very small scales, this process is obviously of such vital
importance that one may hope to find some observable consequences, specific
for particular models of particle physics.
And, indeed, we now believe that there can be some clues left.
Among those are: topological defects production in non-thermal phase
transitions \cite{nth}, GUT scale baryogenesis \cite{bau},
generation of primordial background of stochastic gravitational waves
at high frequencies \cite{gw}, just to mention a few. However, matter
appears in many kinds and forms, and it is hard to review all
possibilities in one talk. I'll concentrate on a possible relation
to a mounting puzzle of the Ultra High Energy Cosmic Rays (UHECR).

When proton (or neutron) propagates in CMB, it gradually looses energy 
colliding with photons and creating pions \cite{gzk}. There is a threshold
energy for the process, so it is effective for very energetic nucleons
only, which leads to the famous Greisen-Zatsepin-Kuzmin (GZK) 
cutoff of the high energy tail of the
spectrum of cosmic rays. All this means that detection of, say,
$3\times 10^{20}$ eV proton would require its source to be within
$\sim 50$ Mpc. However, many events above the cut-off were observed
by Yakutsk, Haverah Park, Fly Eye and AGASA collaborations \cite{cr}
(for the review see Ref. \cite{bs}).

Results from the AGASA experiment \cite{AGASA} are shown in Fig. 1.
The dashed curve represents the expected  spectrum if conventional 
extragalactic sources of UHECR would be distributed uniformly in the 
Universe. This curve displays the theoretical GZK cut-off, but we see events
which are way above it. (Numbers attached to the data points show the 
number of events observed in each energy bin.) Note that no candidate 
astrophysical source, like powerful active galaxy nuclei, were found in 
the directions of all six events with $E > 10^{20}$ eV \cite{AGASA}

\begin{figure}[b!] % fig 1
\centerline{\leavevmode\epsfysize=6.cm \epsfbox{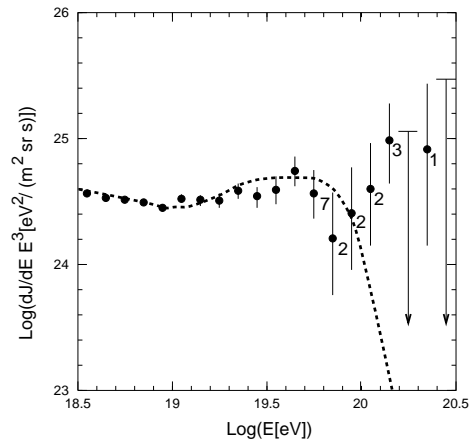}}
%\centerline{\epsfig{fig2_spec_dif.ps,height=3.5in,width=3.5in}}
\vspace{10pt}
\caption{AGASA data set [7], February 1990 -- October 1997.}
\label{fig1}
\end{figure}

There were no conventional explanation found to these observations,
and the question arises, is it indication of the long awayted new physics, 
at last ?

Many solutions to the puzzle were suggested, which rely on different
extensions of the standard model, in one way or the other. Among those
are:
\begin{itemize}
\item A particle which is immune to CMBR. 
In this scenario, primary particle is produced in conventional 
astrophysical accelerators and is able to travel cosmological
distances.  There are variations to this scheme. This can be a new exotic 
particle able to produce normal air showers in Earth's atmosphere 
\cite{farrar}, or this can be an accelerated (anti)neutrino annihilating 
via $Z^{0}$ resonance on the relic neutrinos in a local
high density neutrino clump, thus producing energetic gamma or
nucleon \cite{W}. Massiveness of neutrino, $m_\nu \sim $ eV, is a 
necessary requirement in this scheme.
\item Another possibility is that UHECR are produced when topological
defects destruct near the lab (on the cosmological scale) \cite{hsw}. 
Topological defects which were considered in these kinds of scenarios were:
{strings \cite{S}, }
{superconducting strings \cite{hsw}, }
{networks of monopoles connected by strings \cite{necl}, }
{magnetic monopoles \cite{mon}. }
\item {Conceptually the simplest possibility is that UHECR are produced
(again cosmologically locally) in decays of some new particle\cite{relicX}.}
The candidate $X$-particle must
obviously  obey constraints on mass, number density and lifetime.
\end{itemize}

\section*{UHECR from decaying particles}

In order to produce cosmic rays in the energy range
$E > 10^{11}$ GeV, the decaying primary
particle has to be {\bf heavy}, with the mass well above GZK cut-off,
$m_X > 10^{12}$ GeV. The lifetime, $\tau_{X}$, cannot be much smaller than
the age of the Universe, $\tau_U \approx 10^{10}$~yr. Given this shortest
possible lifetime, the observed flux of UHE cosmic rays will be
generated with the rather low density of $X$-particles,
$\Omega_{X} \sim 10^{-12}$,
where $\Omega_{X} \equiv m_{X} n_{X}/\rho_{\rm crit}$, $n_X$ is the
number density
of X-particles and $\rho_{\rm crit}$ is the critical density.
On the other hand, X-particles must not overclose the Universe,
$\Omega_{X} < 1$.
With $\Omega_{X} \sim 1$, the X-particles may play the role of cold dark
matter and the observed flux of UHE
cosmic rays can be matched if $\tau_{X} \sim 10^{22}$~yr.

The problem of the particle physics mechanism responsible for a long but
finite lifetime of very heavy particles can be solved in several ways.
For example, otherwise conserved quantum number carried by  X-particles
may be broken very weakly due to instanton transitions, or quantum 
gravity (wormhole) effects \cite{relicX}. 
Other interesting models of superheavy long-living particles
were found in Refs. \cite{Xmodels}.

Spectra of UHE cosmic rays arising in decays of relic X-particles
were successfully fitted to the data 
for $m_X$ in the range $10^{12} < m_X/{\rm GeV} < 10^{14}$
\cite{mx}.

Here I address the issue of X-particle abundance.
It was noticed \cite{CKR,KT98} that such heavy particles are produced
in the early Universe from the vacuum fluctuations and their
abundance can be correct naturally, if the standard Friedmann epoch in
the Universe evolution was preceded by the inflationary stage.
This is a fundamental process of particle creation unavoidable in the
time varying background 
and it requires no interactions. Temporal change of the metric is the single
cause of particle production.
Basically, it is the same process which during inflation had generated
primordial large scale density perturbations. 
No coupling (e.g. to the inflaton or plasma) is needed.
All one needs are
stable (very long-living) X-particles with the mass of order of
the inflaton mass, $m_X \approx 10^{13}$ GeV.
Inflationary stage is not required to produce superheavy particles
from the vacuum.
Rather, the inflation provides a cut off in excessive gravitational 
production of heavy
particles which would happen in the Friedmann Universe if it would
start from the initial singularity \cite{KT98}.
Resulting abundance is quite independent of detailed
nature of the particle which makes the superheavy (quasi)stable X-particle
a very interesting dark matter candidate. New particle needs good name.
I like Wimpzilla \cite{wimpZ}.

\paragraph*{Friedmann Cosmology. }

For particles with conformal coupling to gravity (fermions
or scalars with $\xi = 1/6$ in $\xi R\phi^2$ interaction term with 
the curvature), it is the particle mass 
which couples the system to the
background expansion and serves as the source of particle
creation. Therefore, just on dimensional grounds, we expect 
$n_X \propto m_X^3 a^{-3}$ at late times when particle creation diminishes.
In Friedmann cosmology, $a \propto (mt)^\alpha \propto (m/H)^\alpha$
and the anticipated formulae for the X-particles abundance can be 
parameterised as
$n_X = C_\alpha m_X^3 ({H}/{m_X})^{3\alpha}$.
It is expansion of the Universe which 
is responsible for particle creation. Therefore, this equation
which describes simple dilution of already created particles
is valid when already $H << m_X$.
On the other hand particles with $m_X >> H$ cannot be created by this
mechanism. Creation occurs when $H \sim m_X$. Coefficient $C_\alpha$ 
can be found numerically \cite{KT98}, its
typical value is $O(10^{-2})$, and we find that
stable particles with {$m_X > 10^9$ GeV} will overclose the Universe.
There is no room for Superheavy particles in our Universe if it
started from the initial Friedmann singularity \cite{KT98},
since the value of the Hubble constant is limited from above only
by the Planck constant in this case.

\begin{figure}[t!] % fig 2
\centerline{\leavevmode\epsfysize=6.cm \epsfbox{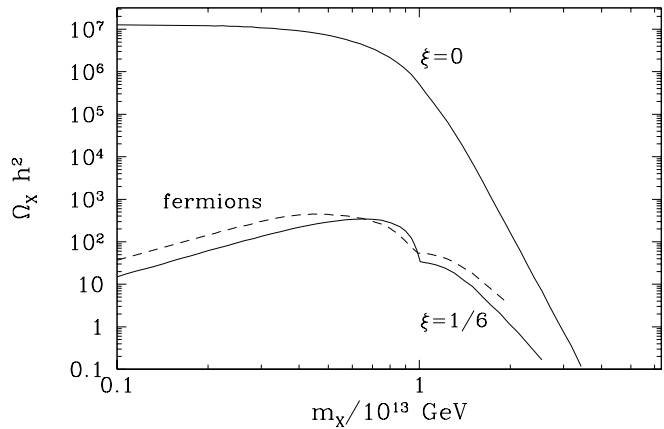}}
\vspace{10pt}
\caption{Ratio of the energy density in $X$-particles, gravitationally 
generated in inflationary cosmology, to the critical energy density is 
shown as a function of X-particle mass, Ref. [18].}
\label{fig2}
\end{figure}

\paragraph*{Inflationary Cosmology. }

If there was inflation, the Hubble constant (in effect) did not exceeded
the inflaton mass, $H < m_\phi$. The mass
of the inflaton field has to be $m_\phi \approx 10^{13}$ GeV as constrained
by the amplitude of primordial density fluctuations relevant for 
the large scale structure formation.
Therefore, production of particles with 
{$m_X > H \sim 10^{13}$ GeV} has to be suppressed in inflationary cosmology.
Results of direct numerical integration of gravitational particle
creation in chaotic inflation model with the potential 
$V(\phi) = m^2_\phi \phi^2/2$ is shown in Fig. 2.

This figure was calculated assuming $T_{\rm R} = 10^9$ GeV for the reheating
temperature. (At reheating the entropy of the Universe was created in 
addition to 
X-particles. In general, multiply this figure by the ratio
$T_{\rm R}/10^9$ GeV and divide it by the fractional entropy increase 
per comoving volume if it was significant at some late epoch.)
Reheating temperature is constrained, $T_{\rm R} < 10^{9}$ GeV, 
in supergravity theory \cite{gtino}. 
We find that $\Omega_X h^2 < 1$ if
$m_X \approx ({\rm few})\times 10^{13}$ GeV.
This value of mass is in the range suitable
for the explanation of UHECR events \cite{KT98}. 
Gravitationally created superheavy 
X-particles can even be the dominating form of matter in the Universe today 
if X-particles are in this mass range \cite{CKR,KT98}. 

\section*{Topological defects and inflation}

\begin{figure}[t!] % fig 3
\centerline{\leavevmode\epsfysize=4.6cm \epsfbox{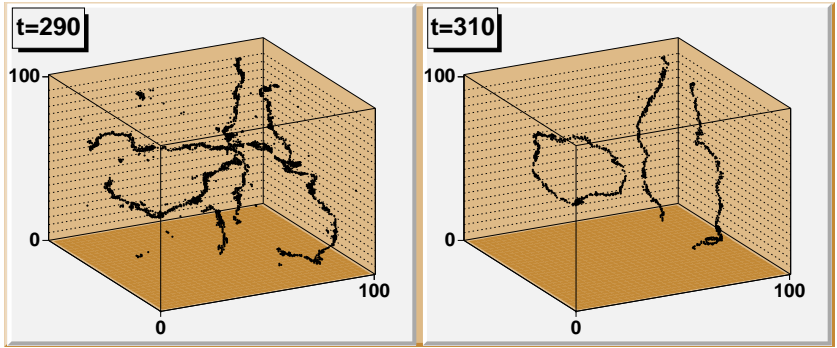}}
\vspace{10pt}
\caption{String distribution at two successive moments of time.}
\label{fig3}
\end{figure}

Decaying topological defect can naturally produce very energetic
particles, and this may be related to UHECR \cite{hsw} -\cite{mon}, 
for recent reviews see \cite{bs}. However, among motivations for
inflation there was the necessity to get rid of unwanted topological defects.
And inflation is excellent doing this job. Since temperature after
reheating is constrained, especially severely in supergravity models, 
it might be that the Universe was never reheated up to the point
of  GUT phase transitions.
Topological defects with a sufficiently high scale of symmetry
breaking cannot be created.
How then topological defects could populate the Universe?

The answer may be provided by non-thermal phase transitions \cite{nth}
which can occur in preheating \cite{preh} after inflation. Explosive
particle production caused by stimulated decay of inflaton oscillations
lead to anomalously high field variances which restore symmetries
of the theory even if actual reheating temperature is small. Defects
form when variances are reduced by the continuing  expansion of the Universe
and phase transition occur.
This problem is complicated, and while some features can be anticipated
and some quantities roughly estimated, the problem requires numerical
study. In recent papers \cite{defects} the defect formation and even
the possibility of the first order phase transitions during preheating
was demonstrated explicitly. Fig.~3 shows string distribution in a simulation
with symmetry breaking scale ${\rm v} =  3 \times 10^{16}$ GeV,
when a pair of ``infinite'' strings and one big loop had formed. 
Size of the box is comparable to the Hubble length at this time.

\section*{Conclusions}

Next generation cosmic ray experiments, which will be soon operational, will
tell us which model for UHECR may be correct and which has to be ruled out.
One unambiguous signature is related to homogeneity and anisotropy of cosmic
rays. If particles immune to CMBR are there, the UHECR events
should point towards distant, extraordinary astrophysical sources \cite{FB98}.
If wimpzillas are in the game, the Galaxy halo will be reflected in anisotropy
of the UHECR flux \cite{DT98}. It is remarkable that we might be able to
learn about the earliest stages of the Universe's evolution. Discovery of heavy
X-particles will mean that the model of inflation is likely correct, or that
at least ``standard'' Friedmann evolution from the singularity is ruled out,
since otherwise X-particles would have been inevitably 
overproduced \cite{KT98}.

\end{document}